\documentclass[journal]{IEEEtran}

\ifCLASSINFOpdf
\else
   \usepackage[dvips]{graphicx}
\fi
\usepackage{url}

\usepackage{amsfonts}

\usepackage{graphicx}

\usepackage{gensymb}

\usepackage{booktabs}

\begin{document} 

\title{A Deep Reinforcement Learning Approach for Audio-based Navigation and Audio Source Localization in Multi-speaker Environments}

\author{Petros Giannakopoulos, Aggelos Pikrakis, and Yannis Cotronis
\thanks{Research in this paper was supported by the Hellenic Foundation for Research and Innovation (H.F.R.I.) under the ``1st Call for H.F.R.I. Research Projects to support Faculty Members \& Researchers and the Procurement of High-Cost Research Equipment Grant'' (Project Number: 3449).}
\thanks{Petros Giannakopoulos is a PhD candidate at the Department of Informatics and Telecommunications, National and Kapodistrian University of Athens, Greece (e-mail: petrosgk@di.uoa.gr).}
\thanks{Aggelos Pikrakis is an Assistant Professor at the Department of Informatics, University of Pireaus, Greece (e-mail: pikrakis@unipi.gr).}
\thanks{Yannis Cotronis is a Professor at the Department of Informatics and Telecommunications, National and Kapodistrian University of Athens, Greece (e-mail: cotronis@di.uoa.gr).}}

\markboth{}
{Shell \MakeLowercase{\textit{et al.}}: Bare Demo of IEEEtran.cls for IEEE Journals}
\maketitle

\begin{abstract}
In this work we apply deep reinforcement learning to the problems of navigating a three-dimensional environment and inferring the locations of human speaker audio sources within, in the case where the only available information is the raw sound from the environment, as a simulated human listener placed in the environment would hear it. For this purpose we create two virtual environments using the Unity game engine, one presenting an audio-based navigation problem and one presenting an audio source localization problem. We also create an autonomous agent based on PPO online reinforcement learning algorithm and attempt to train it to solve these environments. Our experiments show that our agent achieves adequate performance and generalization ability in both environments, measured by quantitative metrics, even when a limited amount of training data are available or the environment parameters shift in ways not encountered during training. We also show that a degree of agent knowledge transfer is possible between the environments. 
\end{abstract}


\IEEEpeerreviewmaketitle

\section{Introduction}

\IEEEPARstart{W}{e} jointly address the problems of real-time audio-based navigation and real-time audio source localization in virtual three-dimensional environments, for the case where the sound sources are human speakers, using an approach based on online Deep Reinforcement Learning. Our objectives are: a) for the problem of audio-based navigation, to investigate whether an autonomous reinforcement learning agent, in the form of an entity able to walk in all directions, can successfully learn to navigate in a three-dimensional space, guiding itself towards a target sound source (a given speaker) while avoiding other sound sources (other speakers), b) for the problem of audio source localization, to investigate whether an autonomous reinforcement learning agent, in the form of a highly directional microphone that is capable of rotating in all directions, can learn to locate sound sources in the surrounding environment and point the microphone towards each detected sound source. In both applications, the agent is trained via a reward-punishment scheme, which is designed to motivate it to solve the environment and discourages it from spending too much time in doing so. In both cases, the only information available to the agent is the raw audio data from the environment, in the form of a two-channel (stereo) audio signal, in an attempt to simulate the  information that a human listener placed inside the environment would receive. The raw audio signal is generated by the Unity game engine to have pseudo-spatial three-dimensional information, meaning that the audio engine employs DSP techniques to simulate a number of channels higher than two and improve directionality of audio sources.

More specifically, for application (a) we created an environment in the Unity game engine \cite{juliani2018unity} which simulates an entity, controlled by the agent, moving in a room where there also exist a number of sound sources in the form of human speakers, placed in random locations in the room. The objective of the entity is to move towards and eventually reach a specified target speaker in the shortest possible amount of time, while at the same time avoiding coming into contact with the other speakers or going outside the room boundaries. Our experiments show that the entity controlled by the deep reinforcement learning agent, rewarded when reaching the target speaker and punished when bumping into other speakers or going outside the room boundaries, is able to solve the aforementioned problem with a degree of success much higher than random chance, i.e., when compared with an agent which moves the entity in a random manner. The agent also outperforms a human in control of the entity. For application (b) the environment consists, again, of a room inside which a number of speakers are placed in random locations. However, in this case, a virtual directional microphone is placed in the center of the room, capable of rotating freely in every direction. An agent is in control of this microphone and can rotate it by applying a corresponding torque to adjust its azimuth and elevation. The objective is to point the microphone towards each speaker in the room until all speakers are detected, in the shortest possible amount of time and with the least possible microphone movement. Our experiments show that the reinforcement learning agent is successful in solving this task, achieving higher detection accuracy than an agent moving the microphone randomly, as well as outperforming a human controlling the microphone.

In both environments, we observed that our reinforcement learning agent is able to generalize well to speakers not seen during training and appears robust to distortion effects not encountered during training, such as various types of sound reverberation.

We also observed that a significant amount of acquired agent knowledge is transferable between the two application environments. The ability to locate sound sources based on environmental audio data was shown to be the most important and difficult skill for the agent to learn. However, once acquired for one application environment, it can be applied to another. The agent, for the most part does not have to re-learn that skill in the new environment if an appropriate transfer learning scheme is used.

The motivation behind this work was to investigate whether it is possible for a reinforcement learning agent to detect the locations and identities of speakers present in its environment and successfully perform tasks based on that information. This has the potential to expand the range of stimulus autonomous agents can use in order to be successful in a variety of environments and tasks. For example, if there are auditory cues in the environment that could be potentially helpful, then an agent capable of incorporating them, possibly along with other types of information (e.g. visual cues) could be more successful than an agent relying only on one type of sensory inputs (e.g. only visual). One practical application could be the development of AI in games which can respond to auditory cues, as well as possibly other stimuli, with a minimal amount of engineering and hard-coded behaviours. Furthermore, it was shown that behaviors acquired by reinforcement learning agents in a simulated environment can, to a degree, transfer to the real world \cite{zhao2020sim}, \cite{rudin2021learning}. Behaviors learned in an appropriately crafted simulated environment, which incorporate auditory sensory information also present in the real world, could potentially be transferable to a real world setting.        

\section{Related Work}

To our knowledge there has been limited published work in the domains of audio-based navigation and audio source localization for methods based on reinforcement learning. The work in \cite{woubie2019autonomous} combined deep reinforcement learning with environmental audio information in the context of navigation of a virtual environment by an autonomous agent. They discovered that an agent trained via reinforcement learning and leveraging raw audio information could reach more reliably a sound-emitting target within a maze compared to the case when only visual information was used. In \cite{hegde2021agents} the authors augmented the VizDoom simulator \cite{kempka2016vizdoom} with the ability to provide audio from the environment to an agent. They showed that an RL agent based on PPO was able to perform a number of simulated tasks requiring audio sensory information. The agent was also able to outperform an agent relying only on visual information in a duel inside the VizDoom simulator. In \cite{lathuiliere2018deep} the authors employed deep reinforcement learning for controlling the gaze of a robotic head based on audio and visual data from a virtual environment. The work of \cite{wang2014sound} focused on development of a framework for sound localization and tracking of sound sources in combination with visual information, for navigation of an autonomous agent in a virtual environment. Their system consists of a sound propagation model and sound localization model based on classical (non-DNN) algorithms. Previous preliminary work by the authors in \cite{giannakopoulos2021deep} created an autonomous agent based on deep reinforcement learning, able to navigate a virtual environment posing an \textit{audio-based navigation} challenge, where a fixed number of speakers are present in a room, and guide itself towards a defined speaker while avoiding colliding with the other speakers, using only raw audio information from the environment. We extend upon this preliminary work by adding another environment focused on the task of \textit{audio source localization}, unifying the training regime of the autonomous agent across both environments and studying the degree of agent knowledge transfer between the environments. 

It has to be noted that supervised deep learning methods have also enjoyed success in recent years over classical methods in the domains of speaker detection and identification \cite{furui200940,sztaho2019deep}, speaker localization \cite{he2018deep,wang2018robust,chakrabarty2017multi} and more general audio source localization tasks \cite{vera2018towards,van2020modelling,yalta2017sound}. Most of these types of approaches largely rely on datasets created from prior recordings of sound sources in specific environments made with special microphone types. 

In our approach we are assessing the viability of methods based on deep reinforcement learning to perform the tasks of detection and localization of human speaker audio sources in three-dimensional spaces, while simultaneously performing continuous control of autonomous entities. We experimentally show that such deep reinforcement learning algorithms have the potential to achieve these tasks while being solely dependent on the auditory sensory information present in the environment. Due to the lack, to our knowledge, of an accepted standard benchmark for reinforcement learning tasks in the audio domain (contrary to the more mature field of reinforcement learning in the visual domain, with standardized benchmarks such as ALE \cite{bellemare2013arcade}) we created our own virtual environments based on the Unity game engine, simulating the tasks described above. We have the ability to customize several parameters about each environment, such as number of sound sources, audio signal parameters of each source, sources placement and realistic audio distortions due to environment characteristics.

\begin{figure}
\begin{minipage}[b]{1.0\linewidth}
  \centering
  \centerline{\includegraphics[width=8.5cm]{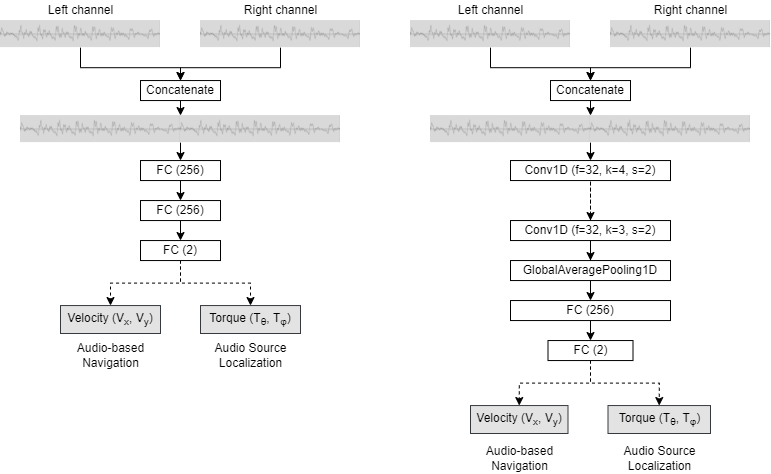}}
\end{minipage}
\caption{Left: Agent's neural network architecture. The left and right channel of a stereo audio signal from the Unity environment are concatenated into a single vector. The resulting vector is then fed into a fully connected feed-forward neural network. The output of the network is either a) the velocity $(V_x, V_y)$ of the agent-controlled entity in the navigation application  or b) the torque $(T_\theta, T_\phi)$  that should be applied to the agent-controlled microphone to adjust its azimuth $\theta$ and elevation $\phi$ respectively in the source localization application. Right: Alternative convolutional neural network architecture, where $f$ is the number of filters, $k$ is the size of the 1-D convolutional kernel, and $s$ is the stride. The features extracted from the concatenated audio input by this convolutional encoder are then averaged and passed to a fully-connected layer. For our experiments we found that this architecture did not provide significant performance benefits over the simpler architecture of only fully-connected layers.}
\label{fig:arch}
\end{figure}

\begin{table}
\caption{Performance comparison of DNN and CNN architectures}
\centering
\begin{tabular}{lcc}
\textbf{Architecture}          & \textbf{Reward}      & \textbf{Success Rate} \\ \midrule
DNN                          & $0.497 \pm 0.053$      & $96\%$                  \\
CNN                        & $0.502 \pm 0.047$      & $96\%$                  \\
\end{tabular}
\label{tab:results_architectures}
\end{table}

\begin{table}
\caption{Performance scaling with number of samples}
\centering
\begin{tabular}{lcc}
\textbf{Number of samples}          & \textbf{Reward}      & \textbf{Success Rate} \\ \midrule
256                          & $0.317 \pm 0.092$      & $91\%$                  \\
512                          & $0.402 \pm 0.069$      & $95\%$                  \\
1024                          & $0.497 \pm 0.053$      & $96\%$                  \\
2048                        & $0.435 \pm 0.072$      & $96\%$                  \\
4096                        & $0.338 \pm 0.098$      & $92\%$                  \\
\end{tabular}
\label{tab:results_num_audio_samples}
\end{table}

\begin{figure}
\begin{minipage}[b]{1.0\linewidth}
  \centering
  \centerline{\includegraphics[width=6.0cm]{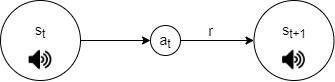}}
\end{minipage}
\caption{Each one of the environments can be described as a Markov Decision Process (MDP), where the state $s$ is the auditory information that an agent is able to extract from the environment at a  time instant $t$. The agent can then perform an action $a$ based on that information and receive a reward $r$ from the environment based on the outcome of its action, before transitioning to the next state.}
\label{fig:mdp}
\end{figure}

\begin{figure}
\begin{minipage}[b]{1.0\linewidth}
  \centering
  \centerline{\includegraphics[width=8.5cm]{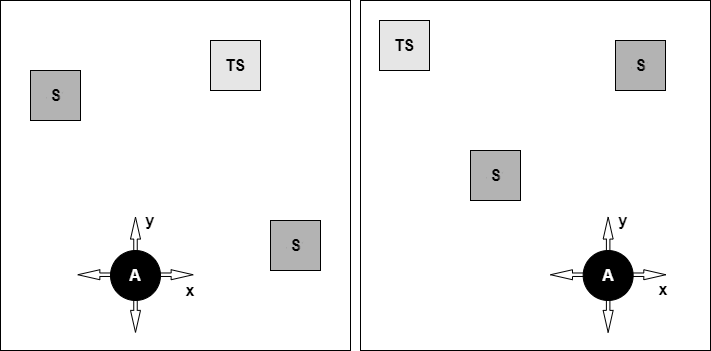}}
\end{minipage}
\caption{Top-down views of the environment used for the audio-based navigation experiments, for the case where $3$ speakers are present. The views show two example state snapshots, one where agent $A$ has direct ``line-of-sight'' to $TS$ (left) and one where it does not (right). Agent $A$ can move along the $x$ and $y$ axes, while the speakers are stationary. $TS$ is the target speaker and $S$ are other speakers in the room whose signals interfere with the signal produced by speaker $TS$. The agent $A$ must reach speaker $TS$ without running into the other speakers $S$.}
\label{fig:navigation_env}

\bigskip

\begin{minipage}[b]{1.0\linewidth}
  \centering
  \centerline{\includegraphics[width=8.5cm]{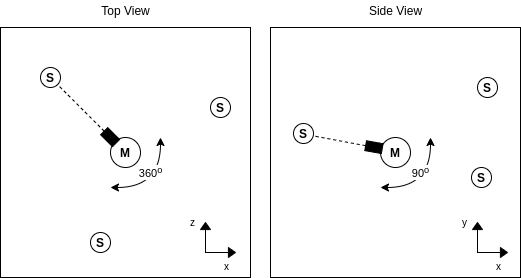}}
\end{minipage}
\caption{Top and Side views of the room where the microphone (M) and speakers (S) are located, for the case where $3$ speakers are present. The microphone is located in the center of the room and it is free to point to any direction by adjusting its azimuth and elevation. Sound sources can be located in a $360$ degrees range around the microphone but they are constrained to be up to $30$ degrees above or below the microphone.}
\label{fig:localization_env}
\end{figure}

\section{Overview of Agent and Environments}

We start by providing detailed descriptions of our methods and the virtual environments that we  created for each one of the aforementioned problems in order to assess our reinforcement learning approach.

\subsection{Agent Architecture}

Our agent architecture is based on a combination of the Proximal Policy Optimization (PPO) online reinforcement learning algorithm \cite{schulman2017proximal} with a small neural network consisting of two fully connected hidden layers. Our choice of reinforcement learning algorithm is not restrictive and other policy gradient optimization algorithms can be potentially suitable as well. The input to the neural network is the left and right channels of the stereo audio signal produced by the Unity game engine, concatenated into a single one-dimensional vector. In our experimental setup we retrieve the most recent $1024$ audio samples from the Unity engine's audio buffer for each channel. This means that, for a sampling frequency of $48$kHz, approximately, the most recent $21$ms  of the audio signal produced by the environment is the input frame to the agent at timestep $t$. Therefore, the resulting vector, after the concatenation operation, has a length of $2048$ samples and it is fed as input to the network (Figure \ref{fig:arch}). Each hidden layer of the network has $256$ neurons. We selected this particular architecture and input scheme after a coarse grid search over different fully-connected and 1D convolutional hidden-layer architectures, lengths of audio buffer and sampling frequencies: 1) In Table \ref{tab:results_architectures} we compare the performance on our audio-based navigation task of the current DNN architecture and a CNN architecture based on a stack of 1D convolutional layers (Figure \ref{fig:conv_arch}). We found that the agent achieves the same performance, within margin of error, with both architectures, however the CNN is more computationally demanding, therefore we decided to use a DNN architecture for the rest of the experiments. However, there is always the possibility that the CNN-based architecture may outperform the simpler DNN-based architecture depending on the environment and task. 2) In Table \ref{tab:results_num_audio_samples} we compare the performance of different lengths of audio buffer given as input to the agent in number of samples. Initially, the agent's performance improves as the length of the audio input increases since there is more audio context available to the agent. However, after a point, performance decreases as the markovian property of the state space required by the reinforcement learning algorithm is violated. Longer audio buffer lengths are more computationally expensive as well. The output of the network is a fully connected layer consisting of $2$ neurons. In the case of audio-based navigation, the network outputs are the estimates for the velocity with which the entity should move in the $x$ and $y$ directions, normalized in the $[-1, 1]$ range. In the case of audio source localization, the network outputs are the estimates of the torque that should be applied to the microphone gimbal in order to adjust the azimuth $\theta$ and elevation $\phi$ of where the microphone should be pointing, also normalized in the $[-1, 1]$ range. The desired outputs are the only difference between the two environments, otherwise the agent's architecture is identical between them.

\subsection{Environments}

In our study, we used the Unity editor to create custom virtual environments that run on the Unity game engine. We chose to create our environments in Unity after investigating various known frameworks and platforms for creating virtual environments suitable for training reinforcement learning agents. The reason we eventually selected Unity was that it provided the easiest to use and most robust set of tools for creating environments containing audio sources and entities with highly customizable properties and also allowed for seamless capturing of audio data from the simulated environment without the need for additional tools.
We present below a detailed description of each the two test environments:

\subsubsection{Audio-based navigation environment}

Our environment focused on audio-based navigation (Figure \ref{fig:navigation_env}) consists of the following key components: a) A rectangular room, b) One up to five speakers (denoted $S$) placed inside the room, one of which is the target speaker (denoted $TS$) and c) The agent ($A$) inside the room. Each speaker is a stationary Unity audio source which plays back an audio file containing an utterance for that speaker, randomly selected from a pool of utterances. When an utterance finishes, another one is randomly selected from the pool and played back. All audio sources share the same properties and are accordingly configured in the Unity editor. Specifically, the signal intensity (volume) of the audio source decreases linearly  with the distance from the source (linear volume roll-off). The maximum distance from which an audio source can still be heard is set so that all sources are audible across the entire room. The agent ($A$) can move inside the room along the $x$ and $y$ axes by adjusting its velocity vectors $v_x$ and $v_y$ over each axis. Each training episode begins with the agent appearing in a random location of the lower edge of the room and the speakers show up in random locations inside the room. The audio sources are placed at various heights above the floor of the room in order to simulate speakers of different heights, e.g. one speaker can be $1.70$m tall while another may be $1.85$m tall and their mouths are therefore located at different heights. The agent receives a positive reward of $+1.0$ when it reaches speaker $TS$ and a negative reward of $-1.0$ when it crashes onto another speaker $S$ or when it moves outside the room boundaries. It also receives a small negative reward of $-0.001$ for every step it takes to discourage the agent from procrastinating. A training episode ends  when the agent reaches the target speaker, one of the other speakers or moves outside the room boundaries.

\subsubsection{Audio source localization environment}

The environment focused on audio source localization (Figure \ref{fig:localization_env}) consists of three key components: a) A rectangular room, b) One up to five speakers, $S$, placed inside the room, c) A highly directional microphone, $M$, placed on an omnidirectional gimbal able to freely rotate in every direction. The microphone apparatus is placed in the center of the room. The agent can adjust the direction that the microphone ($M$) is pointing to by applying the appropriate amount of torque to it, in order to change its azimuth ($\theta$) and elevation ($\phi$) angles. Each training episode begins with the microphone at its default starting rotation of $\theta = 0\degree$ and $\phi = 0\degree$. A random number of sound sources, ranging from $1$ to $5$ on each episode, are placed at random locations inside the room. Each source represents a speaker located at a certain distance from the microphone and at a certain height. Like in the audio-based navigation environment, each speaker is an audio source playing back a randomly selected utterance, audible across the entire room with a linear volume roll-off relative to the distance from the source. The agent in control of the microphone receives a positive reward of $+1.0$ when it has pointed it towards all of the speakers and a small negative reward of $-0.001$ for every action it takes, in order to encourage it to complete the task as soon as possible. A training episode ends either after the agent has pointed the microphone towards every speaker or when a predefined time limit has ran out.

For the sake of reproducibility, the Unity project files have been made publicly available over the Internet \footnote{https://github.com/petrosgk/AudioRL}, including all required components for running the experiments, i.e., the Unity environments, the pre-trained agents and the datasets.

\section{Training the Agent}

Each environment can be formalized as a Markov Decision Process (Figure \ref{fig:mdp}) with the following characteristics: a) The state, $s_t$, at time $t$ consists of  two 1-D vectors, representing the left and right channels of the raw audio waveform of the environment, which are subsequently concatenated into a single 1-D vector, b) Given $s_t$, the agent takes an action $a_t$, which consists of a $1$-D vector made up from two scalars. For the audio-based navigation environment, each scalar is representing, the normalized velocity of the moving entity on the horizontal and vertical axis ($V_x$, $V_y$), respectively, with $V_x \in [-1, 1]$ and $V_y \in [-1, 1]$. For the audio source localization environment, each scalar is representing the torque ($T_\theta$, $T_\phi$), with $T_\theta \in [-1, 1]$ and $T_\phi \in [-1, 1]$, which are applied to adjust the azimuth and elevation of the agent-controlled microphone, respectively, c) As a result of action $a_t$, the agent transitions to the next state $s_{t+1}$ and receives a reward $r \in [-1, 1]$.

We assume that there exists an optimal policy $\pi^*$, which, when followed by the agent, maximizes the cumulative reward, $r$, achieved for a horizon of $T$ time steps, where each timestep is an interaction with the environment. The agent's training objective is to find a close approximation, $\pi_\theta \approx \pi^*$, to the optimal policy $\pi_\theta$, where $\theta$ are the parameters of a neural network. To train the network parameters, the policy runs for $T$ time steps and the collected samples are used for updating the policy gradient. In order to do this we first need to estimate the advantage at timestep $t$ of action $a_t$ in state $s_t$, as it was formulated in \cite{mnih2016asynchronous}:

\begin{equation}
    A_t(a_t, s_t) = \delta_t + \gamma\delta_{t+1} + \dots + \gamma^{T-t+1}\delta_{T-1},
\end{equation}
with
\begin{equation}
    \delta_t = r_t + \gamma V^{\pi}(s_{t+1}) - V^{\pi}(s_t),
\end{equation}

\begin{figure}
\begin{minipage}[b]{1.0\linewidth}
  \centering
  \centerline{\includegraphics[width=8.5cm]{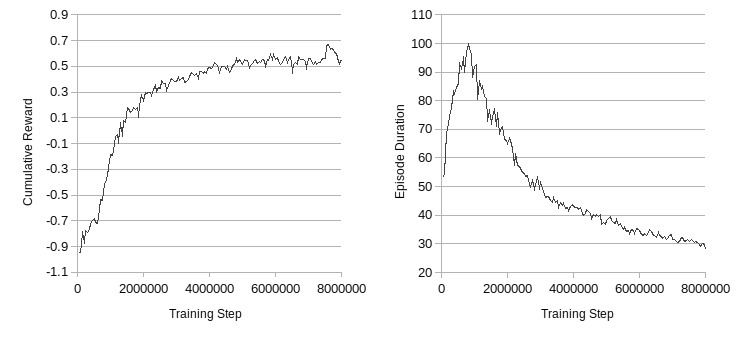}}
\end{minipage}
\caption{Left: Progression, in terms of cumulative reward achieved in an episode, during training the agent in the audio-based navigation environment. Right: Average number of steps that the agent needed to complete an episode as training proceeded. The number of steps starts low in the beginning of training as the agent tends to go outside the room boundaries or crashes into other speakers, then rises sharply as the agent tends to procrastinate, then drops sharply while the reward rises as the agent learns to reach the target speaker.}
\label{fig:training_curves_navigation}
\end{figure}

\begin{figure}
\begin{minipage}[b]{1.0\linewidth}
  \centering
  \centerline{\includegraphics[width=8.5cm]{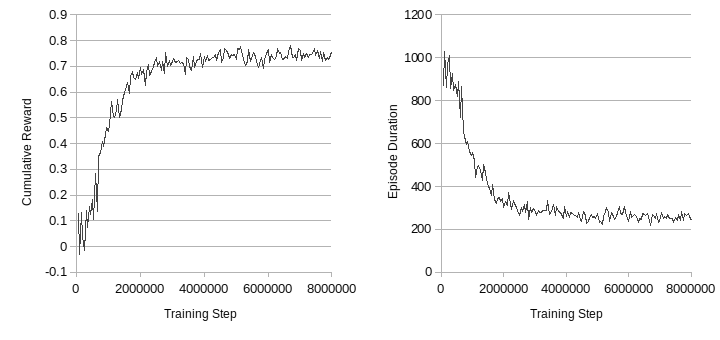}}
\end{minipage}
\caption{Left: Progression, in terms of cumulative reward achieved in an episode, during training  the agent in the audio source localization environment. Right: Average number of steps that the agent needed to complete an episode as training proceeded. The number of steps decreases as the agent learns to locate all the speakers present in the room faster.}
\label{fig:training_curves_localization}
\end{figure}

where
\begin{equation}
    V^{\pi}(s) = \mathbb{E}[\sum_{k=0}^\infty\gamma^kr_{t+k}]
\end{equation}
is the value estimate of state $s$ under policy $\pi$ and it is the mean expected reward, over $k$ time steps, for following policy $\pi$ from state $s$.
In the above, $t \in [0,T]$ is the timestep index, $r_t$ is the reward at timestep $t$ and $\gamma \in (0, 1]$ is the future reward discount factor. The objective minimized by PPO is, according to [14]:
\begin{equation}
    L_t(\theta) = \mathbb{E}_t[L_t^{C}(\theta) - c_1 L_t^{VF}(\theta) + c_2 S[\pi_{\theta}](s_t)],
\end{equation}
where
\begin{equation}
    L^{C}(\theta) = \mathbb{E}_t[min(r_t(\theta)A_t, clip(r_t(\theta), 1-\epsilon, 1+\epsilon)A_t)]
\end{equation}
and
\begin{equation}
   L_t^{VF}(\theta) = (V_\theta(s_t) - V_t^{target})^2
\end{equation}
$c_1$, $c_2$ are coefficients, $S$ is an entropy bonus added to policy estimation, which encourages exploration, and $\epsilon$ is a hyper-parameter. In all our experiments we set $\gamma = 0.99$, $c_1 = 0.95$, $c_2 = 0.001$ and $\epsilon = 0.2$.

\section{Experiments}

In this section we outline our testing methodology and present our experimental results for each of the two environments.

\subsection{Testing Methodology}

The datasets used for the experiments for each environment were created from $10$ publicly available audiobooks, each read by a different speaker. We first split each audiobook in utterances in an automated way, using a Voice Activity Detector (VAD) which labels the boundaries of each utterance. We then manually verified the VAD results and corrected the boundaries of detected utterances where needed. We assume that the definition of an utterance is the linguistic one: \textit{"An uninterrupted chain of spoken or written language."} \cite{lexico_dictionaries}. This procedure resulted into a dataset of approximately $600$ utterances for each one of the speakers. 

\subsubsection{Audio-based navigation environment}

For the audio-based navigation environment, we pick $5$ speakers from the dataset. We then randomly split the pool of utterances for each one of the $5$ speakers into a training partition and a testing partition. The resulting training set consists of approximately $500$ utterances for each speaker, while the test set consists of around $100$ utterances per speaker. We then set one speaker as the target and train the agent for a total of $8$ million steps in the created  environment. Figure \ref{fig:training_curves_navigation} shows: a) the progression of the agent's performance during training in terms of the cumulative reward achieved during an episode relative to the number of training steps the agent has taken in the environment and b) the number of steps it took to complete an episode, by reaching the target speaker, another speaker, or going outside the room boundaries, relative to the number of training steps. Upon completion of the training stage, we evaluate the agent's performance by switching to the test set and letting the trained agent play $100$ episodes in the environment. For each episode we keep a record of the cumulative reward achieved by the agent, which is a quantitative measure of its performance. We mentioned in our description of the environment that the agent receives a large positive reward when it reaches the target speaker, a large negative reward when it collides with another speaker or goes outside the room boundaries, and a small negative reward for each step it takes. Therefore, the higher the cumulative reward, the quicker the agent has reached the target speaker without running into another speaker or going outside the room. After the $100$ testing episodes are completed, we compute the average cumulative reward achieved over all testing episodes and then compare it with the corresponding reward achieved by a baseline agent performing a random action policy as well as a human in control of the agent. The human player was first allowed to familiarize himself with the environment and the task. He can move the agent using the arrow keys on the keyboard and has to complete the task using only auditory information (e.g. without looking at the screen). This training-testing experiment is repeated five times by setting each time one of the five speakers as the target speaker. In the end, we average the cumulative reward  over all target speakers. We also calculate the overall success rate (successfully reaching the target speaker) and likewise compare it with the success rate of the random and human baselines.

\begin{table}
\caption{Results for Audio-based Navigation Environment}
\centering
\begin{tabular}{lcc}
\textbf{Agent Type}          & \textbf{Reward}      & \textbf{Success Rate} \\ \midrule
PPO                          & $0.497 \pm 0.053$      & $96\%$                  \\
Human                        & $-0.910 \pm 0.082$      & $16\%$                  \\
Random                       & $-1.052 \pm 0.042$     & $4\%$                   \\
                             & \multicolumn{1}{l}{} & \multicolumn{1}{l}{}  \\
\textbf{Training Utterances} & \textbf{Reward}      & \textbf{Success Rate} \\ \midrule
500                          & $0.497 \pm 0.053$      & $96\%$                  \\
50                           & $0.348 \pm 0.046$      & $89\%$                 \\
1                            & $0.274 \pm 0.046$      & $67\%$                 \\
                             & \multicolumn{1}{l}{} & \multicolumn{1}{l}{}  \\
\textbf{Reverberation Type}  & \textbf{Reward}      & \textbf{Success Rate} \\ \midrule
No reverberation             & $0.497 \pm 0.053$      & $96\%$                  \\
Room (low reverberation)     & $0.391 \pm 0.052$      & $93\%$                 \\
Auditorium (high reverberation)     & $0.418 \pm 0.063$      & $92\%$       \\
                             & \multicolumn{1}{l}{} & \multicolumn{1}{l}{}  \\
\textbf{Pitch Shift}  & \textbf{Reward}      & \textbf{Success Rate} \\ \midrule
No pitch shift             & $0.497 \pm 0.053$      & $96\%$                  \\
Pitch shift     & $0.329 \pm 0.077$      & $91\%$                 \\
\end{tabular}
\label{tab:results_navigation}

\bigskip

\caption{Results for Audio Source Localization Environment}
\centering
\begin{tabular}{lcc}
\textbf{Agent Type}          & \textbf{Reward}      & \textbf{Success Rate} \\ \midrule
PPO                          & $0.748 \pm 0.021$      & $74\%$                  \\
Human                        & $0.105\pm 0.095$      & $10\%$                  \\
Random                       & $-0.158 \pm 0.082$     & $2\%$                   \\
                             & \multicolumn{1}{l}{} & \multicolumn{1}{l}{}  \\
\textbf{Training Utterances} & \textbf{Reward}      & \textbf{Success Rate} \\ \midrule
500                          & $0.748 \pm 0.021$      & $74\%$                  \\
50                          & $0.621 \pm 0.055$      & $70\%$                  \\
1                            & $0.427 \pm 0.039$      & $59\%$                 \\
                             & \multicolumn{1}{l}{} & \multicolumn{1}{l}{}  \\
\textbf{Reverberation Type}  & \textbf{Reward}      & \textbf{Success Rate} \\ \midrule
No reverberation                           & $0.748 \pm 0.021$      & $74\%$    \\
Room (low reverberation)                   & $0.691 \pm 0.049$      & $71\%$   \\      
Auditorium (high reverberation)                   & $0.688 \pm 0.034$      & $70\%$   \\
                             & \multicolumn{1}{l}{} & \multicolumn{1}{l}{}  \\
\textbf{Pitch Shift}  & \textbf{Reward}      & \textbf{Success Rate} \\ \midrule
No pitch shift             & $0.748 \pm 0.021$      & $74\%$                  \\
Pitch shift     & $0.682 \pm 0.083$      & $71\%$ 
\end{tabular}
\label{tab:results_localization}
\end{table}

\subsubsection{Audio source localization environment}

For the audio source localization environment, we pick all $10$ speakers from the dataset. We then split them into a training set and a test set, each containing $5$ speakers. The training set contains $3$ male and $2$ females speakers, while the test set contains $2$ male and $3$ female speakers. Each speaker is unique to the training and test set. We subsequently train the agent for a total of $8$ million steps in the created environment. Figure \ref{fig:training_curves_localization} shows: a) the progression of the agent's performance during training in terms of the cumulative reward achieved during an episode relative to the number of training steps the agent has taken in the environment and b) the number of steps it took to complete an episode by finding all speakers in the room relative to the number of training steps. Upon completion of the training stage, we evaluate the agent's performance similarly to the audio-based navigation environment, i.e., we compute the average cumulative reward and success rate over $100$ testing episodes and then compare them with a random baseline agent and a human controller. Similar to the audio-based navigation environment, the human player can rotate the microphone using the arrow keys on the keyboard and has to complete the task without visual information. In this case, the agent receives a large positive reward when it finds all speakers in the room and a small negative reward for each step it takes. Therefore, the higher the cumulative reward, the quicker the agent has pointed the microphone towards all speakers in the room. The agent's success rate is defined in this environment as having located all speakers in the room before the predefined time limit has lapsed.

For both application environments, we also evaluate the agent's ability to learn when only a small amount of training data is available. To that end, during training, we limit the number of utterances per speaker to $50$ and $1$ (from $500$). In this way, we can observe if the agent can still learn when just a few training data are available per speaker.

Furthermore, we also evaluate the agent's ability to generalize to environment variables that it did not encounter during training. To that end, we add reverberation to the sound sources, which is a step towards bringing the virtual environments closer to realism, since in reality the speakers voices would bounce off the walls, floor and ceiling of the rooms. We try out two levels of reverberation, low and high, corresponding to different environments. We only add reverberation in the testing phase, i.e., reverberation is not present during training. Additionally, we randomly shift the pitch frequency of each speaker by $4$ to $8$ percent, during test time only, and compute again the agent's success rate.

\subsection{Results}

\subsubsection{Audio-based navigation environment}

Our findings for the audio-based navigation environment are summarized in Table \ref{tab:results_navigation} for the three aforementioned experiments. In more detail, the table shows: 
\renewcommand{\labelenumi}{\alph{enumi})}
\begin{enumerate}
\item The average cumulative reward and success rate achieved over $100$ episodes by our agent based on PPO, compared to the average cumulative reward and success rate achieved by an agent randomly navigating the room and by a human player: This result shows that our agent is able to navigate the room and reach the target speaker in a short amount of time, with performance higher than with random chance and a human player. This can be interpreted as an indication that our proposed RL agent is capable of learning a representation of the environment that leads it to identify the target speaker and its position with a high degree of accuracy, based only on raw audio data. We observed that, while a human is quickly able to infer the approximate locations of the speakers, they have trouble doing so with enough precision relative to their position to effectively navigate the environment, unless also using visual information. 
\item The reward achieved by our agent when trained on the full training set of $500$ utterances per speaker, when it is only trained on $50$ utterances per speaker and when it is trained on only $1$ utterance per speaker: Like before, the agent is again evaluated on the same test set of $100$ utterances per speaker. When trained on the small training set of $50$ utterances per speaker we observe a drop of approximately $30\%$ in the reward achieved and a $7\%$ drop in success rate. This result shows that, while the agent now struggles more to reach the target speaker, as evidenced by the lower reward, it can still reach the target speaker with a high degree of success. When trained on the absolute minimum training set of just $1$ utterance per speaker, the achieved reward drops by $45\%$ and the success rate drops by $30\%$. While the drop in success rate is now significant, the agent still outperforms the human and random baselines. We can interpret these results as an indication that the agent can generalize satisfactorily from very little training data. 
\item The reward achieved by our agent when there is reverberation present in the environment during the testing phase but not during the training phase. The cumulative reward achieved by the agent suffers on average a $20\%$ drop with both types of reverberation, compared to the case when no reverberation is present, while the success rate drops by $4\%$ on average. We can conclude from this result that the agent is not particularly affected by distortions due to reverberation, even if those are not encountered during training.
\item The reward achieved by our agent when the pitch of the speakers is shifted during the testing phase. The pitch remains unaltered during the training phase. The cumulative reward achieved by the agent when the pitch is shifted is lower by $10\%$, while the success rate drops is $5\%$. We can conclude that the agent is mostly robust to changes in speakers pitch, even if those were not encountered during training. 
\end{enumerate}

\subsubsection{Audio source localization environment}

Our findings for the audio source localization environment are summarized in Table \ref{tab:results_localization}, similarly to the results for the audio-based navigation environment. We record the average cumulative reward and success rate achieved by the agent over $100$ played episodes after: 
\renewcommand{\labelenumi}{\alph{enumi})}
\begin{enumerate}
\item  training the agent on the full training set,
\item training the agent on a reduced training set, 
\item including reverberation distortions in the environment at test time,
\item introducing random shifts in the speakers pitch at test time.
\end{enumerate}
Then we compare with the performance achieved by random and human baselines. We can observe that the performance of the agent far outperforms by far both a random and a human baseline. It is also able to generalize well from a small amount of training data, suffering a modest $4\%$ drop in success rate when trained on the small training set of $50$ utterances per speaker, while suffering a significant $15\%$ drop in success rate when trained on the absolute minimum training set of $1$ utterance per speaker. The agent also appears to be robust to different types of reverberation, added only during the testing phase, suffering a modest $3-4\%$ drop in success rate depending on the amount of reverberation. The agent also does not appear to be significantly affected by random changes in the speakers pitch, added only during the testing phase, since its success rate then drops by $3\%$. Similarly to the audio-based navigation application, the poor human performance can be explained by the fact that, while a human was easily able to infer the general direction of incoming audio, they were not able to precisely pin-point it, unlike the PPO agent, unless also using visual information.

\begin{figure}
\begin{minipage}[b]{1.0\linewidth}
  \centering
  \centerline{\includegraphics[width=8.5cm]{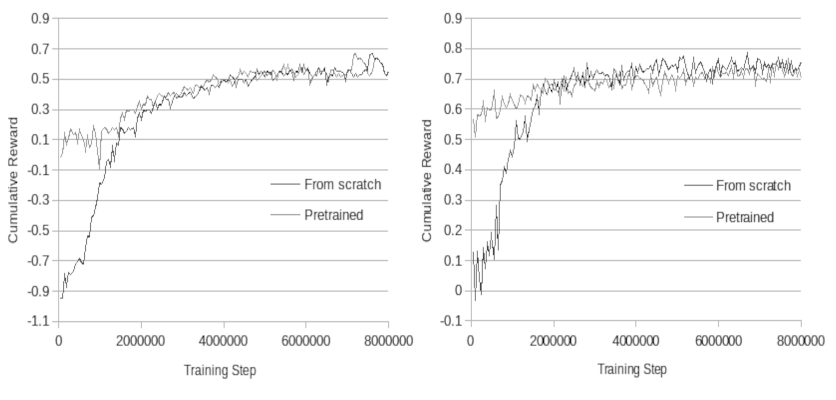}}
\end{minipage}
\caption{Left: Training progression on the \textit{audio-based navigation} environment of: a) an agent trained from scratch on the environment, b) an agent that was pre-trained on the \textit{audio source localization} environment. Right: Training progression on the \textit{audio source localization} environment of: a) an agent trained from scratch on the environment, b) an agent that was pre-trained on the \textit{audio-based navigation} environment. We can observe that the ability to do audio source localization is the most important and difficult skill to learn in both environments.}
\label{fig:transfer_learning}
\end{figure}

\subsection{Transfer Learning}

We investigate whether there exists a degree of transferable skills between the \textit{audio-based navigation} and \textit{audio source localization} tasks, meaning that an agent trained on one task might be able to perform better on the other task compared to an agent who has not encountered either task before. For example, in the \textit{audio-based navigation} environment it is plausible to assume that the agent also implicitly learns to perform audio source localization as part of tackling the navigation task. We can make this assumption if we consider that, in order for the agent to successfully navigate the environment and find the target speaker, it would need to  acquire implicit knowledge about the locations of the speakers in the room from the environmental audio data. Then it must distill that knowledge down to the information it deems necessary to accomplish its task. To test this hypothesis, we first took an agent trained on the \textit{audio source localization} task and then fine-tuned it on the \textit{audio-based navigation} task. We observed that the pre-trained agent starts at a higher performance level, reaches the final performance of an agent trained from scratch in less training steps, but does not exceed it. Similarly, we started with an agent trained on the \textit{audio-based navigation} task and fine-tuned it on the \textit{audio source localization} task. We again observed that the pre-trained agent performs better and reaches peak performance much faster than an agent starting from scratch. Both experiments are illustrated in Figure \ref{fig:transfer_learning}. An agent trained on the \textit{audio source localization} task and then evaluated on the \textit{audio-based navigation} task, without any further training on this task, achieves a success rate of $72\%$ compared to $96\%$ when trained on this task (Table \ref{tab:results_transfer_learning_navigation}). An agent trained on the \textit{audio-based navigation} task and then evaluated on the \textit{audio source localization} task achieves a success rate of $61\%$ compared to $74\%$ when trained on this task specifically (Table \ref{tab:results_transfer_learning_localization}). These results show that the agent learns to perform implicit audio source localization as part of the \textit{audio-based navigation} task. Furthermore, it appears that the ability to do audio source localization is the most important component of the \textit{audio-based navigation} task as well, while the navigation ability component appears easier for the agent to learn. We can conclude that the skill of \textit{audio source localization} is what the agent spends most of its time acquiring, as evident by the degree of this knowledge transfer between the \textit{audio source localization} and \textit{audio-based navigation} environments.

\begin{table}
\caption{Transfer Learning Results for Navigation Environment}
\centering
\begin{tabular}{lcc}
\textbf{Trained for}          & \textbf{Reward}      & \textbf{Success Rate} \\ \midrule
Navigation                          & $0.497 \pm 0.053$      & $96\%$        \\
Localization                        & $0.135 \pm 0.049$      & $72\%$       \\
\end{tabular}
\label{tab:results_transfer_learning_navigation}

\bigskip

\caption{Transfer Learning Results for Localization Environment}
\centering
\begin{tabular}{lcc}
\textbf{Trained for}          & \textbf{Reward}      & \textbf{Success Rate} \\ \midrule
Localization                          & $0.748 \pm 0.021$      & $74\%$        \\
Navigation                        & $0.558 \pm 0.074$      & $61\%$       \\
\end{tabular}
\label{tab:results_transfer_learning_localization}
\end{table}

\section{Conclusions}

We examined the viability of applying deep reinforcement learning on the problems of real-time \textit{audio-based navigation} and \textit{audio source localization} in three-dimensional spaces. For this purpose we designed and constructed, using the Unity game engine, two virtual environments, representative of the above tasks. We also built an agent based on the PPO online reinforcement learning algorithm which accepts raw stereophonic environmental audio data as the state representation at a given time instant and maps it to actions that should be performed given this representation. After training the agent in each  environment, we found that it is able to outperform both a random and a human baseline. Furthermore, the agent was still able to achieve satisfactory performance when trained on less training data. It also showed robustness to distortions due to different types of sound reverberation not encountered during training, suffering a small or zero performance impact. The agent was also robust to random changes in the pitch of the speakers voice, which were not encountered during training. Finally, we examined the degree with which skills learned by the agent in one application environment can be transferred to the other. We concluded that the \textit{audio source localization} skill learned in the respective environment can be, to a large degree, transferred to the \textit{audio-based navigation} task, reducing the number of time steps that the agent must spend to learn the new environment and reach peak performance. We also concluded that \textit{audio source localization} is the most important and difficult skill that the agent needs to acquire in order to succeed in both environments.

\bibliographystyle{IEEEtran}
\bibliography{main}

\end{document}